# ACADEMIC FREEDOM AND INNOVATION: A RESEARCH NOTE

**Authors:** D. Audretsch[1], C. Fisch[2], C. Franzoni[3], P. P. Momtaz[4,5,6]*, S. Vismara[7]

**Affiliations:**

[1] O'Neill School of Public & Environmental Affairs, Indiana University, Bloomington 47405, USA

[2] Interdisciplinary Centre for Security, Reliability and Trust (SnT), University of Luxembourg, 29, Avenue John F. Kennedy, 1855 Luxembourg, Luxembourg.

[3] School of Management, Polytechnic University of Milan, Piazza Leonardo da Vinci 32, 20133 Milan Italy.

[4] University of California, Los Angeles (UCLA), Anderson School of Management, 110 Westwood Plaza, CA 90095, Los Angeles, United States.

[5] TUM School of Management, Arcisstr. 21, 80333 München, Germany.

[6] University College London, Computer Science, Centre for Blockchain Technologies, Malet Place, London, United Kingdom.

[7] University of Bergamo, Department of Management, Via dei Caniana, 2, 24127 Bergamo, Italy.

* Correspondence to: momtaz@ucla.edu.

# ACADEMIC FREEDOM AND INNOVATION: A RESEARCH NOTE

**Abstract:**

The first-ever article published in *Research Policy* was Casimir's (1971) advocacy of academic freedom in light of the industry's increasing influence on research in universities. Half a century later, the literature attests to the dearth of work on the role of academic freedom for innovation. To fill this gap, we employ instrumental variable techniques to identify the impact of academic freedom on the quantity (patent applications) and quality (patent citations) of innovation output. The empirical evidence suggests that improving academic freedom by one standard deviation increases patent applications and forward citations by 41% and 29%, respectively. The results hold in a representative sample of 157 countries over the 1900–2015 period. This research note is also an alarming plea to policymakers: Global academic freedom has declined over the past decade for the first time in the last century. Our estimates suggest that the decline of academic freedom has resulted in a global loss quantifiable with at least 4.0% fewer patents filed and 5.9% fewer patent citations.

1. **INTRODUCTION**

The very first article published in *Research Policy* was Casimir's (1971) work on academic freedom (pages 3–8 in Volume 1 Issue 1). In his precursor to the vast literature on private-public R&D partnerships, Casimir (1971) proposed that the collaboration between industry organizations and universities can only be beneficial for society as long as academic freedom is not at risk. Freedom, in general, has been recognized as a source of growth since the beginning of modern economic thought (Smith, 1937). Free-market societies develop faster because freedom spurs innovation (Schumpeter, 1942). Institutions that allow for free cooperation and competition promote knowledge production and exploitation (Acemoglu and Robinson, 2012; North, 1992). Independence from authority and hierarchy fosters an atmosphere of information exchange and tolerance to failures that spurs idea circulation, experimentation, diversity, and creativity; all ultimately inducing innovation (Florida, 2014; Aghion et al., 2008). Academic freedom is similarly embedded in the norms of science, along with other cornerstones of science, such as open disclosure and freedom of critique (Merton, 1973; Polanyi, 1947). These norms are key to promoting the kind of unconditional exploration that fuels science and would not be possible in the private sector, because highly uncertain economic returns and limited appropriability discourage profit-oriented researchers (Nelson, 1959; Arrow, 1962; Dasgupta and David, 1994).

Half a century after Casimir's article, the relevance of academic freedom for innovation has never been empirically tested. Reopening the debate on academic freedom and innovation seems especially important at the present days, given that academic freedom is arguably under threat (Behrens and O'Grey, 2001; Nelson, 2004) and indicators have registered signs of a decline in the last decade, for the first time in the post-WWII period (Spannagel and Kinzelbach, 2022).

To address this gap, we compile a comprehensive country-year panel including 157 countries over the 1900–2015 period. We measure innovation output both in terms of *quantity*, proxied by the aggregate number of patent applications in a given country and year, and *quality*, proxied by the aggregate number of forward citations after three years to the aggregate number of patent applications in a given country and year. Annual data on country-level academic freedom come from the V-Dem Institute at the University of Gothenburg. Overall, our sample covers up to 12,392 country-year observations that account for 62.8M patent applications and 36.8M citations, equivalent to a sample population coverage ratio of 92%, and therefore well-representative of the inventive activity since the beginning of the 20$^{th}$ century.

Figure 1 presents a strikingly strong association between academic freedom and the quantity (i.e., patent applications; Figure 1, top) and quality (i.e., forward citations; Figure 1, bottom) of innovation output. A one standard deviation improvement in academic freedom corresponds to 300 more patent filings and 50 more forward citations after five years per one million inhabitants.

**[PLEASE INSERT FIGURE 1 ABOUT HERE]**

However, estimating the impact of academic freedom on innovation at the country level requires precluding other potential reasons for the association observed in Figure 1. One potentially confounding explanation relates to spurious correlation. For example, rather than academic freedom, it might be that perceived freedom, correlated with both academic freedom and innovation output, is driving the association. Another potentially confounding explanation is reverse causality. That is, a country might adapt its level of academic freedom in response to an insufficient amount of past or contemporaneous innovation output. To overcome these threats of endogeneity, we implement a battery of instrumental variable tests. These approaches instrument current academic freedom with the level or stock of previous academic freedom or liberal democracy to address reverse causality and spurious correlation concerns, respectively. Instrumenting current academic freedom with past academic freedom addresses reverse causality concerns if the lag is chosen for a long-enough period. Instrumenting academic freedom with liberal democracy isolates the variation in academic freedom from the variation in general freedom and therefore uses as an estimator only the part of academic freedom that stems from *academic* freedom, addressing concerns about the perceived level of general freedom as a spurious confounder. Taken together, these instruments, if yielding consistent results, can be interpreted as collective evidence that the association in Figure 1 is likely causal.

Regardless of the empirical strategy chosen to identify causality, our results consistently suggest that academic freedom has strong positive effects on both the quantity and the quality of innovation output. First, an improvement in (uninstrumented) academic freedom by one standard deviation increases the number of patent applications two years later by 41%. The estimated effects for the instrumented academic freedom range from 37% to 61%, indicating that the estimated effect is robust. In fact, Montiel et al.'s (2013) weak-instrument tests suggest that all four instruments (level and stock of past academic freedom and liberal democracy) are good instruments in a statistical sense. Second, an improvement in uninstrumented academic freedom by one standard deviation increases the number of forward citations five years later by 29%, and instrumented academic freedom yields effects of even larger magnitude. In post-hoc sensitivity tests, we show that these effects are robust to (i) the inclusion of additional country-level controls, (ii) different lags with which our dependent variables are measured (i.e., patent applications and forward citations), (iii) testing the models in highly innovative versus non-innovative countries, (iv) testing the post-1980 period only to account for country-year panel imbalance, and (v) using an alternative, *de jure* measure of academic freedom.

We close by commenting that the positive relationship found between academic freedom and innovation also means that the decline in academic freedom registered in the latest years could hinder the innovation rate and prosperity of countries in the years to come.

## 2. BACKGROUND

### *2.1 Academic freedom: Concept and country heterogeneity*

Academic freedom is "the right to choose one's own problem for investigation, to conduct research free from any outside control, and to teach one's subject in the light of one's own opinions" (Polanyi, 1947).

Although ancient Greek philosopher Socrates (who preferred to be sentenced to death rather than accepting any constraint to his freedom of expression of his philosophical thought) is credited in Plato's "The Trial and Death of Socrates" for being the first human to assert academic freedom, the institutionalization of the modern right of academic freedom took until the early 20th century (Stone, 2015). In the US, the focus of post-Civil War higher education shifted from preparing students for clergy and elite professions to training them for practical jobs, which resulted in researchers and teachers demanding independence from governing bodies. This led to the founding of the American Association of University Professors (AAUP) in 1915, which, together with the American Association of Colleges and Universities, published the Statement of Principles on Academic Freedom and Tenure in 1940. The AAUP describes academic freedom as "the freedom of a teacher or researcher in higher education to investigate and discuss the issues in his or her academic field and to teach or publish findings without interference from political figures, boards of trustees, donors, or other entities. Academic freedom also protects the right of a faculty member to speak freely when participating in institutional governance, as well as to speak freely as a citizen".

After World War II, the importance of academic freedom was reinstated, not only as a way to insure freedom of opinion from the influence of the political governments, but also as an ingredient necessary to unleash the full creative potential of science (Stone, 2015). For example, in 1946, Oppenheimer, quoting Fermi, declared that "an intensive freedom of the individual scientific worker [..] is the only way to insure that no important line of attack is neglected" (US Senate, 1945).

At present, the right to academic freedom is (legally) protected with considerable heterogeneity across jurisdictions. In the US, it is linked with the constitutional right of expression and speech in the First Amendment since the Supreme Court case of Sweezy v. New Hampshire (1959). However, the constitutional protection of the concept of academic freedom is limited to public universities, not to private institutions; pursuant to the State Action Doctrine and the First Amendment's right of freedom of expression and speech, it does not cover all realms of academic freedom under the AAUP's definition, such as the right to determine class curricula. In the EU, the right of academic freedom is paragraphed in Article 13 of the Charter of Fundamental Rights, which states that the "arts and scientific research shall be free of constraint. Academic freedom shall be respected." However, there is no common definition of academic freedom and member states have discretion over the enforcement of the right (Karran, 2007).

## 2.2 Threats to academic freedom

There is considerable evidence that academic freedom is under serious threat recently both in democratic and non-democratic countries. The Scholars at Risk (2021) report lists 332 incidences of severe violations of academic freedom, including 110 cases of killings, violence, and disappearances, 101 cases of imprisonment and 34 cases of loss of position. Violators are "state and non-state actors, including armed militant and extremist groups, police and military forces, government authorities, off-campus groups, and even members of higher education communities". Beyond the serious incidents occurred mostly under repressive regimes, in the last decade academic freedom has been threatened also in several countries traditionally seen as guarantors of academic freedom.

Threats to academic freedom have come from three main sources. First, interest groups moved by moral, religious, or ideological agendas voice concerns and attack professors on social media for their opinions, research, or teaching in controversial areas (Reichman, 2019). Examples include stem cell research, the use of animal models, and, more recently, views on the COVID-19 pandemic (Nogrady, 2021). Second, some countries have regulations that allow political power to exert direct control over universities. For example, in China, all department chairs and deans are centrally-appointed, and a regulation that mandates a Party-appointed leader in all departments has been enforced since 2013 (Acemoglu et al., 2022). Finally, in some countries, the pursuit of for-profit opportunities in higher education institutions is progressively shifting the governance structures of academia, away from the collegiate model and toward a managerial model, which is more akin to a corporate research culture (Nogrady, 2021). This may induce academics to conform to institutional priorities and to eschew research themes that may be disliked by powerful donors and constituents, thus constraining research exploration.

## 2.3 The recent decline in global academic freedom

Research has documented a concerning decline in academic freedom at the global level in the last decade, for the first time since WWII (Spannagel and Kinzelbach, 2022). For the purpose of the present research, we focus on the subset of the 25 countries with the strongest science base, identified as the top-20 countries of the Scimago Country Ranking (1996-2021, all subject areas) either by H-Index, or by citations received.[1] Figure 2 reports the level of academic freedom registered in the 25 countries in 2011 and 2021. The graph shows a large heterogeneity across countries in 2021. Countries in Europe and North America, Australia, and South Korea have average AFI above 0.75. China, Iran, and Saudi Arabia have average AFI below 0.25. The comparisons of the mean country values in 2021 and 2011 indicate that only one country, South Korea, registered an improvement (+0.07) in the last decade. Fourteen countries remained overall stable (variations smaller than 0.02). Ten countries registered a reduction in academic freedom greater than -0.02. The most dramatic decreases were registered in Brazil (-0.56), Turkey (-0.43), India (-0.39) and Russia (-0.25). The USA (-0.15), UK (-0.13), and China (-0.12) also experienced a decrease. Over a longer time-span, the average level of AFI across the 25 countries improved from 1941 to 2001, plateaued-off around 2001 to 2007, and declined afterward. The level of AFI in 2021 (0.70) is approximately equivalent to the one registered by the same countries in 1985.

**[PLEASE INSERT FIGURE 2 ABOUT HERE]**

---

[1] The source of country ranking data was accessed on October 3rd, 2022 at: https://www.scimagojr.com/countryrank.php.

## 2.4  Conceptual background: (Academic) freedom and innovation

The general concept of freedom has been recognized as a source of growth since the beginning of modern economic thought (Smith, 1937). Neoclassical economics argues that free-market societies develop faster because freedom spurs innovation (Schumpeter, 1942). Also, institutional economics studies institutional design choices that enable free cooperation and competition to promote the production and exploitation of knowledge (North, 1992; Acemoglu and Robinson, 2012). Knowledge and innovation initiate spillover effects, which is a reason for why freedom and innovation could be mutually reinforcing (Audretsch and Feldman, 1996; Audretsch and Belitski, 2013; Acs et al., 2012). For example, the network theory of social capital argues that social freedom attracts talented people and promotes an atmosphere of tolerance, which fosters information exchange, knowledge spillovers, diversity and creativity, all ultimately inducing innovative productivity (Florida, 2014).

To understand why academic freedom may relate to innovation, we must first explain why academia is important for innovation. Governments subsidize academics to invent because if inventing were left to the private sector alone, society would underinvest in innovative projects (Nelson, 1959; Arrow, 1962). One reason is that, given the existence of geographically-bounded knowledge spillovers and the absence of perfect intellectual property rights, private agents would not be able to fully appropriate their inventions' economic value.

The type of R&D that is best done by academia vis-á-vis industry illustrates why academic freedom plausibly spurs innovation. The optimal division of inventive effort between academia and industry is early-stage and late-stage R&D, respectively. One reason is that early-stage R&D is highly uncertain and therefore benefits from the inventors' creative freedom to find a solution, while late-stage R&D is less uncertain and requires directed focus on monetizing an invention which is best achieved with hierarchy and delegated authority in the corporate setting (Aghion et al., 2008). Another reason is that the open exchange of ideas in academia facilitates knowledge spillovers but creates the risk of ideas being stolen. Instead the control over information exchanges that occurs in corporate settings impedes knowledge spillovers and protects ideas from being stolen (Hellmann and Perotti, 2011).

Both arguments build on the notion of academic freedom. Academic freedom is a necessary prerequisite for both creativity and open information exchange (Shaw, 2022; Ekboir, 2003; Casimir, 1971). Moreover, academia has various idiosyncratic norms, rules, and incentives that rely on academic freedom and that distinguish academic from corporate inventive efforts (Behrens and Gray, 2001; Lee, 1996). Academic institutional features, such as the tenure system, peer review, and rewards for impactful work, promote the production and diffusion of knowledge (Partha and David, 1994; Franzoni and Rossi-Lamastra, 2017). These institutional features work effectively only with a sufficiently high degree of academic freedom.

## 3. EMPIRICAL DESIGN AND RESULTS

### 3.1 Data

#### 3.1.1 Data sources

We compile a country-level panel dataset for 157 countries over the 1900–2015 period, merging data from three different sources. Innovation-related data come from PATSTAT, the most comprehensive database on worldwide patent activity maintained by the European Patent Office.[2] Academic freedom data, as well as data to instrument academic freedom, come from the 2022 release of the Academic Freedom Dataset from the V-Dem Institute of the University of Gothenburg[3], the primary and most credentialled international source providing an explicit measurement of the concept of academic freedom (Spannagel and Kinzelbach, 2022). Finally, data for country-level control variables are provided by the World Bank.

#### 3.1.2 Variables

We study the effect of changes in academic freedom on innovation output, as represented by patents. We measure *innovation quantity* by the number of patent applications filed in a country/year and *innovation quality* by the number of citations received by the patents of a country/year in the first 3 years after issuance. The final dataset comprises information on 62.8M patent applications and 36.8M citations related to the 157 countries for which the AFI is available in the time window 1900–2015. This corresponds to a sample/population coverage of 92% in PATSTAT.

The Academic Freedom Index (AFI) from the University of Gothenburg's V-Dem Institute is a country/year metric of academic freedom provided in the 0–1 range, for 175 countries over more than 120 years (Spannagel and Kinzelbach, 2022). The AFI is obtained by aggregating the independent opinions of more than 2,000 country experts, who are usually academics based in the countries. The experts are asked to consider de facto states of five key indicators: freedom to research and teach, freedom of academic exchange and dissemination, institutional autonomy of universities, campus integrity (intended as freedom from surveillance and harassment in campuses and digital learning platforms), and freedom of academic and cultural expression (Spannagel and Kinzelbach, 2022). The independent expert opinions are calibrated and aggregated by point estimates drawn from a Bayesian factor analysis aimed at minimizing potential biases due to multiple-raters and issues of consistency across countries and years (Coppedge et al., 2021; Pemstein et al., 2022).

---

[2] https://www.epo.org/searching-for-patents/business/patstat.html. Accessed September 13, 2022.
[3] https://www.v-dem.net/our-work/research-programs/academic-freedom/. Accessed January 30, 2023.

Our baseline estimation controls for country- and year-fixed effects to capture all time-variant and between-country confounding factors. In additional tests, we show our results are robust to various modifications to our baseline estimation, such as the inclusion of additional controls, such as GDP per capita, population size, and migration rate.

### 3.1.3 Summary statistics

Summary statistics are reported in Table 1. The number of patent applications (natural logarithm, leaded by two years) is on average 1.908 with a standard deviation of 3.206. Thus, the average country of our 157 countries over the 1900–2015 period files 6.7 patents per year with a standard deviation of 24.7 patents. On average, these patents receive 1.6 forward citations over the next three years with a standard deviation of 5.1. There is substantial positive skewness in these variables, given that the minimum is at 0 and the maximum is at around 1.3 million for both the number of patent application and forward citations, highlighting the need for country-fixed effects in our empirical models. The academic freedom index is z-standardized (mean = 0, standard deviation = 1). Not reported in Table 1, the average country-year observation in our sample has a GDP per capita of $8,655, a population of 7.5 million inhabitants, and a migration rate of 52.6%.

**[PLEASE INSERT TABLE 1 ABOUT HERE]**

### 2.2  Identification strategy to mitigate endogeneity concerns

Our baseline models regress the two country-year innovation output proxies (aggregate patent applications and aggregate forward citations) on the AFI, and country- and year-fixed effects. This specification may incur two specific concerns about its ability to identify whether any estimated association between academic freedom and innovation output is causal. First, a concern with the baseline specification is that there might be other, omitted factors, such as the level of general freedom in a given country, that are correlated with both academic freedom and innovation output (spurious correlation). Second, another concern is that a country may adjust its level of academic freedom in response to the state of the country's current innovation output (reverse causality). Specifically, instead of academic freedom driving innovation output, the concern is that innovation output (or a relative lack thereof) may drive academic freedom.

To address these concerns, we implement four distinct instrumental variable approaches from the political analysis literature. First, we instrument academic freedom with the country's five-year-lagged liberal democracy level (approach 1) and with the country's liberal democracy stock calculated over the previous ten years with a 1% deprecation rate (approach 2). Intuitively, approaches 1 and 2 address concerns about spurious correlations, e.g., that the general level of freedom confounds the measured effect of academic freedom on the innovative performance in a country, and instrument academic freedom with an index of country-year liberal democracy, following Edgell et al. (2020). Second, we instrument academic freedom with the country's five-year-lagged AFI, assuming that reverse causality concerns do not exist for more than that period (approach 3), and with the country's academic freedom stock calculated over the previous ten years (approach 4). Approaches 3 and 4 use values of the AFI in previous years as an instrument to address concerns of reverse causality, i.e., that the inventiveness of a country induces more academic freedom. Moreover, weak-instrument tests based on Montiel et al. (2013) for all four approaches indicate the goodness of our instruments in statistical terms.

### 3.3 *Main results: Academic freedom and the quantity and quality of innovation output*

Figure 1 established the stylized fact that, for the average country in our sample over the 1900–2015 period, a one standard deviation improvement in academic freedom is associated with 300 more patent filings and 50 more forward citations after three years per one million inhabitants. Here, we conduct uninstrumented and instrumented fixed-effects panel regressions to test whether the association is causal.

Table 2 shows regression results for our innovation *quantity* proxy. The dependent variable is the natural logarithm of the two-year-leaded aggregate number of patent applications per year and country. The independent variable in Model 1 is the z-standardized academic freedom index per country and year. The independent variables in Models 2, 3, 4, and 5 are the instrumented versions of the AFI using the country's five-year-lagged liberal democracy level, liberal democracy stock calculated over the previous ten years with a 1% deprecation rate, five-year-lagged AFI, and academic freedom stock calculated over the previous ten years, respectively. The number of country-year observations ranges from 10,753 to 12,392 across model specifications, with the discrepancy due to limited availabilities of the instruments for some country-years. The (adjusted) R-squared in all of our models exceeds 80%, indicating that our panel-fixed-effects specification captures more than four-fifth of the variation in innovation output in terms of aggregate country-level patent applications. Finally, instrument statistics show that lagged and stock liberal democracy and lagged and stock academic freedom are strong instruments. The Montiel et al. (2013) weak-instrument test always rejects the null that our instruments of academic freedom are "weak" at the 5% significance level.

The regression results across all models show a positive and statistically significant effect of academic freedom on patent applications, which suggests that an increase (decrease) in academic freedom leads to an increase (decrease) in the quantity of innovation output. The baseline (uninstrumented) estimation in Model (1) yields a coefficient of the z-standardized AFI of 0.345 (robust standard error = 0.032), with p-value <1%, suggesting that a one standard deviation increase in academic freedom increases the average number of patent applications by 41.2% (=exp(0.345)-1). The 2SLS tests in Models (2) to (5) are consistent with the uninstrumented result in Model (1), indicating that academic freedom has a positive effect on the number of patent applications. For all four instrumented versions of academic freedom, we estimate coefficients in the range of 0.315 and 0.477, with p-values always below 1%. Therefore, depending on the instrument, improving academic freedom by one standard deviation translates into 37% to 61% more patent applications two years later.

**[PLEASE INSERT TABLE 2 ABOUT HERE]**

Table 3 shows regression results for our innovation *quality* proxy. The dependent variable is the natural logarithm of the five year-leaded aggregate number of forward citations on the aggregate number of granted patents per year and country. The independent variables are the same as those in Table 2. The only difference next to the replacement of the dependent variable is the inclusion of the natural logarithm of the number of patent applications as a control variable. Conditioning on the number of patent applications helps ensure that any identified effect on the quality of innovation output is indeed due to academic freedom and not through a mediating mechanism (i.e., academic freedom increases patent applications and therefore also forward citations). Hence, the coefficients on academic freedom in Table 3 can be interpreted as the per-patent impact on forward citations of improving academic freedom.

The baseline regression estimates a coefficient of 0.258 (robust standard error = 0.022), with a p-value less than 1% in Model (1). Therefore, improving academic freedom by one standard deviation translates into more forward citations per patent in the amount of 29%. The four instrumented versions suggest that academic freedom's causal effect on forward citations is even higher. The estimated coefficients in the 2 SLS specifications range from 0.431 to 1.149, statistically significant always at the 1% level. Note that our fixed-effects panel model captures are large amounts of variation in the dependent variable, with R-squares always larger than 60%. Further, the Montiel et al. (2013) weak-instrument test again rejects the null hypothesis that our four instruments are weak at the 5% level.

**[PLEASE INSERT TABLE 3 ABOUT HERE]**

*3.4    Robustness checks*

We estimate a series of additional models to assess the robustness of our results. The results are reported in Tables 4 and 5 for the quantity and quality of innovation output, respectively. Panels A in Tables 4 and 5 displays the effect of academic freedom on innovation output with additional control variables (i.e., GDP growth, population (ln), migration rate). The main effect of AFI on innovation output remains positive and significant across all specifications. In our main analyses, our dependent variable innovation output is forwarded by two years to consider a publication lag of patent applications. The models in Panels B of Tables 4 and 5 consider different lags of the dependent variable and show that the main results are robust when considering innovation output forwarded by 0 to 4 years. Panels C in Tables 4 and 5 consider the effect of academic freedom on innovation output with the sample split according to the total number of patent applications per country. The results show that the effect of academic freedom on innovation output is pronounced in countries with a high number of patent applications. The effect is less pronounced when only considering countries that file fewer patent applications (Model 4). Model 5 shows that the positive association holds when excluding countries without any patenting activity. In a similar vein, Panels D in Tables 4 and 5 only consider observations after 1980 because most of the patenting activity in our sample occurred since then. Again, the main results hold and remain unchanged in significance and magnitude. Last, Panels E of Tables 4 and 5 use a different measure for academic freedom. Specifically, they consider a measure of de jure academic freedom (summary statistics: mean = 0.361, standard deviation = 0.480) that captures whether academic freedom was constitutionally protected in a given country in a given year. Like our de facto measure of academic freedom, the de jure measure comes from the V-Dem Institute's Academic Freedom dataset. The positive association between academic freedom and innovation output holds when the de jure measure is considered.

**[PLEASE INSERT TABLES 4 AND 5 ABOUT HERE]**

## 4. DISCUSSION AND CONCLUDING REMARKS

Does academic freedom impact innovation? Although the question has attracted a good deal of attention in the research-policy literature, the aggregate impact of academic freedom on both the quantity and quality of patenting activity around the world is underexplored (e.g., Ekboir, 2003; Lee, 1996). Motivated by this research gap, we leverage a novel data source, the Academic Freedom Index, to investigate the relation between academic freedom and innovation in a representative sample including 157 countries over the 1900-2015 period. Employing instrumental variable methods to identify the causal effect of academic freedom on innovation output, we find that academic freedom positively affects the number of patent applications and the per-patent number of forward citations. According to our baseline model estimates, improving academic freedom by one standard deviation increases the number of patents two years later by 41% and the number of per-patent forward citations five years later by 29%. The results are robust to several sensitivity tests.

To the best of our knowledge, this is the first study that explicitly tests the impact of academic freedom on the quantity and quality of patenting activity. Understanding the role of academic freedom for innovation informs the broader policy debates about promoting science (Shaw, 2022; Goldstein and Kearney, 2020; Behrens and Gray, 2001; Lee, 1996; Casimir, 1971). Our study provides empirical evidence on the importance of ensuring academic freedom for policymakers and other stakeholders engaged in "science diplomacy" (Rüffin, 2020, p. 82; Ekboir, 2003). This is important for corporate R&D managers as it informs them about a crucial aspect of the geographical location choice of their R&D activities (e.g., Albino-Pimentel et al., 2022).

We present the relation between academic freedom and innovation as a brief research note with the double purpose to stimulate a debate among policymakers and spur future research to deepen our understanding of this stylized fact. For policymakers, our note's conclusion in combination with the recent decline in academic freedom should be alarming. Academic freedom had progressively increased from the 1940s to the 2010s, but it reversed and started to decline in the last decade both at the global level and in the 25 leading countries in science (see Figure 2). Based on our estimates, the global decline in academic freedom that occurred in the last decade has resulted in a global loss quantifiable in the range of 4.0 to 6.7% fewer patents filed and 5.9 to 23.5% fewer patent citations. For researchers, the newly available data on academic freedom across countries paves the way for further investigating the contingency factors of when and how a lack of academic freedom impedes innovation. In this regard, while our brief research note's purpose was to establish the academic freedom-innovation relation, future research may explore the various channels through which policymakers may improve academic freedom to promote patenting activity.

**Exhibits**

**Figure 1:** Quantity (top) and quality (bottom) of innovation as a function of academic freedom

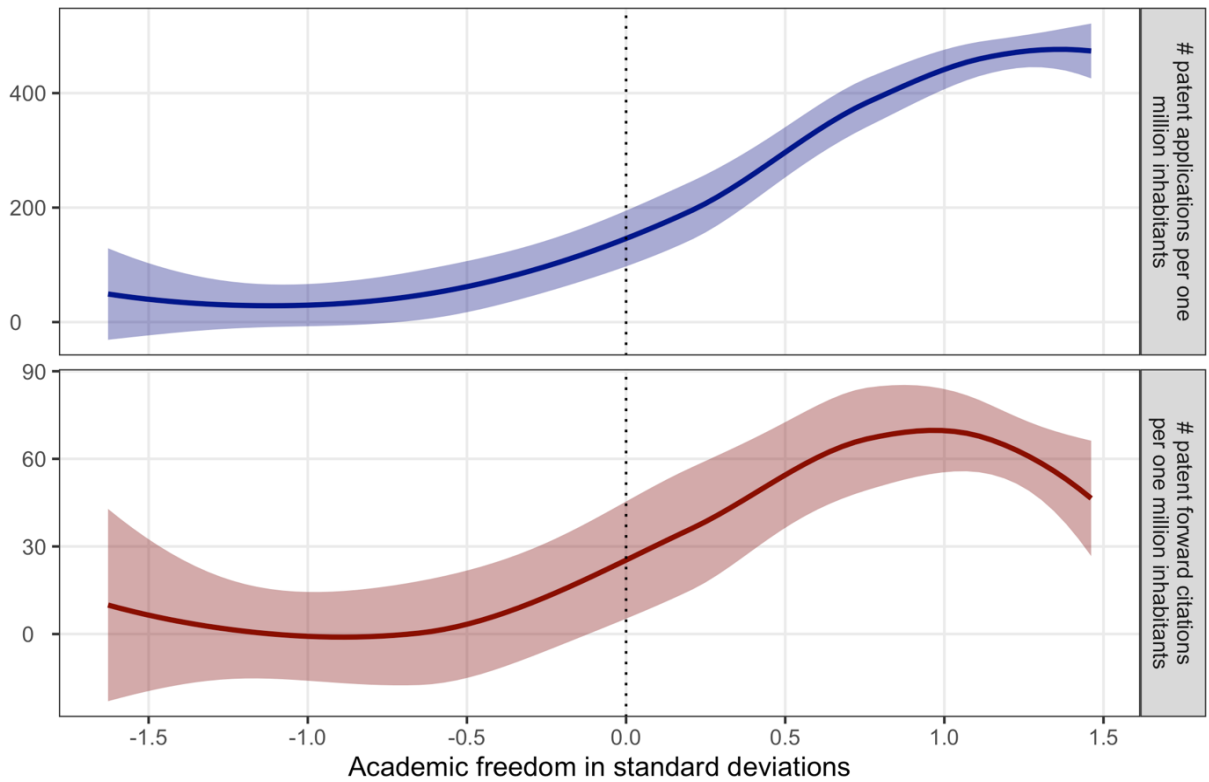

*Note:* Lines display the coefficients of a LOESS regression without additional control variables. Red and blue areas represent 95% confidence intervals.

**Figure 2:** Academic Freedom in 2011 and 2021 in 25 leading countries by science base.

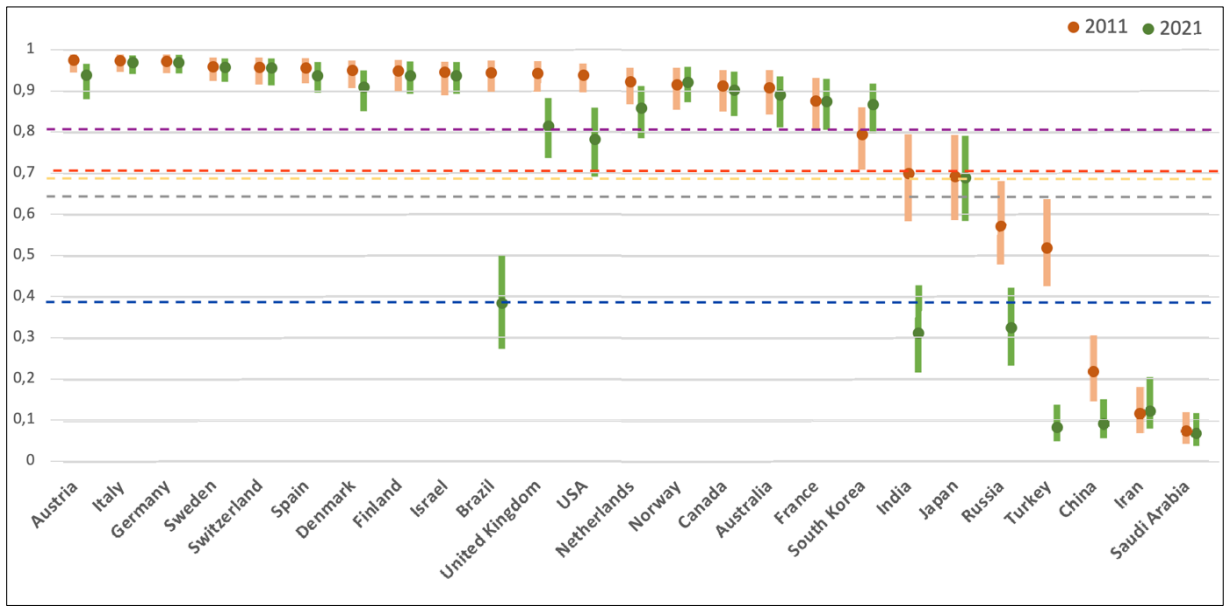

Countries ordered in descending order of Academic Freedom Index (AFI) in 2011. Interval bars represent min-max. Blue dashed line: average country level of AFI in 1941 (0.38). Gray dashed line: average country level of AFI in 1961 (0.65). Yellow dashed line: average country level of AFI in 1981 (0.69). Purple dashed line: average country level of AFI in 2001 (0.80). Red dashed line: average country level of AFI in 2021 (0.70).

**Table 1:** Academic freedom (per year) and innovation activity (per year) across countries.

| Variable | Obs. | Mean | Std. dev. | Min. | Max. |
|---|---|---|---|---|---|
| ***Dependent variables:*** | | | | | |
| Patent applications$_{t+2}$ (ln) | 18,212 | 1.908 | 3.206 | 0.000 | 14.138 |
| Forward citations$_{t+5}$ (ln) | 17,741 | 0.440 | 1.633 | 0.000 | 14.129 |
| | | | | | |
| **Independent variable:** | | | | | |
| Academic freedom$_t$ (z) | 12,706 | 0.000 | 1.000 | -1.620 | 1.476 |

**Table 2:** Main analysis of the effect of academic freedom (AF) on innovation quantity (patent applications).

| Model | (1) | (2) | (3) | (4) | (5) |
|---|---|---|---|---|---|
| **Dependent variable** | ln(Patent applications$_{t+2}$) | ln(Patent applications$_{t+2}$) | ln(Patent applications$_{t+2}$) | ln(Patent applications$_{t+2}$) | ln(Patent applications$_{t+2}$) |
| **Independent variables** | Coeff. (SE) | Coeff. (SE) | Coeff. (SE) | Coeff. (SE) | Coeff. (SE) |
| AF$_t$ (z) | 0.345 (0.032)*** | | | | |
| AF$_t$ (IV: Democracy$_{t-5}$, z) | | 0.477 (0.060)*** | | | |
| AF$_t$ (IV: Democracy stock$_{t-10}$, z) | | | 0.442 (0.057)*** | | |
| AF$_t$ (IV: AF$_{t-5}$, z) | | | | 0.330 (0.043)*** | |
| AF$_t$ (IV: AF stock$_{t-10}$, z) | | | | | 0.315 (0.042)*** |
| | | | | | |
| Year fixed-effects | Yes | Yes | Yes | Yes | Yes |
| Country fixed effects | Yes | Yes | Yes | Yes | Yes |
| | | | | | |
| Countries (observations) | 157 (12,392) | 157 (11,809) | 157 (11,257) | 157 (11,569) | 156 (10,753) |
| R² (R² adjusted) | 0.811 (0.807) | 0.815 (0.811) | 0.818 (0.814) | 0.817 (0.812) | 0.820 (0.816) |
| MOP weak instrument test (F) | - | 2,395.13 ($\tau < 0.05$) | 4,114.46 ($\tau < 0.05$) | 5,946.18 ($\tau < 0.05$) | 11,895.90 ($\tau < 0.05$) |

*Notes*: The estimates in Model (1) are obtained from an OLS regression. The estimates in Models (2)–(5) come from 2SLS regressions. Models (2) and (3) use democracy as an instrument for academic freedom. Democracy is lagged by 5 years in Model (2). Model (3) uses the stock of democracy over the previous 10 years with a deprecation rate of 1%. Models (4) and (5) use values of the academic freedom index in previous years as an instrument. Robust standard errors in parentheses. The MOP weak instrument is based on Montiel Olea and Pflueger (2013) as implemented in STATA's WEAKIVTEST command. High F-values indicate that the instrument is strong. * $p < 0.10$, ** $p < 0.05$, *** $p < 0.01$.

**Table 3:** Main analysis of the effect of academic freedom (AF) on innovation quality (forward citations received in the first 3 years).

| Model | (1) | | (2) | | (3) | | (4) | | (5) | |
|---|---|---|---|---|---|---|---|---|---|---|
| **Dependent variable** | $\ln(\text{Forward citations}_{t+5})$ | | $\ln(\text{Forward citations}_{t+5})$ | | $\ln(\text{Forward citations}_{t+5})$ | | $\ln(\text{Forward citations}_{t+5})$ | | $\ln(\text{Forward citations}_{t+5})$ | |
| **Independent variables** | Coeff. | (SE) | Coeff. | (SE) | Coeff. | (SE) | Coeff. | (SE) | Coeff. | (SE) |
| $AF_t$ (z) | 0.258 | (0.022)*** | | | | | | | | |
| $AF_t$ (IV: Democracy$_{t-5}$, z) | | | 1.149 | (0.057)*** | | | | | | |
| $AF_t$ (IV: Democracy stock$_{t-10}$, z) | | | | | 1.126 | (0.053)*** | | | | |
| $AF_t$ (IV: $AF_{t-5}$, z) | | | | | | | 0.431 | (0.031)*** | | |
| $AF_t$ (IV: AF stock$_{t-10}$, z) | | | | | | | | | 0.436 | (0.031)*** |
| | | | | | | | | | | |
| ln(Patent applications) | Yes | | Yes | | Yes | | Yes | | Yes | |
| Year fixed-effects | Yes | | Yes | | Yes | | Yes | | Yes | |
| Country fixed effects | Yes | | Yes | | Yes | | Yes | | Yes | |
| | | | | | | | | | | |
| Countries (observations) | 157 | (11,921) | 157 | (11,338) | 157 | (10,786) | 156 | (11,099) | 156 | (10,285) |
| $R^2$ ($R^2$ adjusted) | 0.642 | (0.633) | 0.607 | (0.598) | 0.631 | (0.621) | 0.659 | (0.650) | 0.680 | (0.671) |
| MOP weak instrument test (F) | - | | 2,114.24 | ($\tau < 0.05$) | 3,610.46 | ($\tau < 0.05$) | 5,313.93 | ($\tau < 0.05$) | 10,607.19 | ($\tau < 0.05$) |

*Notes*: The estimates in Model (1) are obtained from an OLS regression. The estimates in Models (2)–(5) come from 2SLS regressions. Models (2) and (3) use democracy as an instrument for academic freedom. Democracy is lagged by 5 years in Model (2). Model (3) uses the stock of democracy over the previous 10 years with a deprecation rate of 1%. Models (4) and (5) use values of the academic freedom index in previous years as an instrument. Robust standard errors in parentheses. The MOP weak instrument is based on Montiel Olea and Pflueger (2013) as implemented in STATA's WEAKIVTEST command. High F-values indicate that the instrument is strong. * $p < 0.10$, ** $p < 0.05$, *** $p < 0.01$.

**Table 4:** Robustness tests for the effect of academic freedom (AF) on innovation quantity (patent applications).

**Panel A:** The effect of academic freedom (AF) on innovation output (patent applications) <u>with additional control variables</u>.

| Model | (1) | | (2) | | (3) | | (4) | | (5) | |
|---|---|---|---|---|---|---|---|---|---|---|
| **Dependent variable** | $\ln(\text{Patent applications}_{t+2})$ | | $\ln(\text{Patent applications}_{t+2})$ | | $\ln(\text{Patent applications}_{t+2})$ | | $\ln(\text{Patent applications}_{t+2})$ | | $\ln(\text{Patent applications}_{t+2})$ | |
| **Independent variables** | Coeff. | (SE) | Coeff. | Coeff. | Coeff. | (SE) | Coeff. | (SE) | Coeff. | (SE) |
| $AF_t$ (z) | 0.275 | (0.037)*** | | | | | | | | |
| $AF_t$ (IV: Dem.$_{t-5}$, z) | | | 0.221 | (0.070)*** | | | | | | |
| $AF_t$ (IV: Dem. stock$_{t-10}$, z) | | | | | 0.250 | (0.065)*** | | | | |
| $AF_t$ (IV: $AF_{t-5}$, z) | | | | | | | 0.316 | (0.053)*** | | |
| $AF_t$ (IV: AF stock$_{t-10}$, z) | | | | | | | | | 0.305 | (0.052)*** |
| *Controls* | | | | | | | | | | |
| GDP per capita | -0.005 | (0.004) | -0.004 | (0.004) | -0.004 | (0.004) | -0.005 | (0.005) | -0.007 | (0.005) |
| Ln(Population) | 0.112 | (0.113) | 0.148 | (0.124) | 0.166 | (0.125) | 0.216 | (0.119)* | 0.301 | (0.125)** |
| Migration rate | 0.001 | (0.001)* | 0.001 | (0.001)* | 0.001 | (0.001)* | 0.001 | (0.001)* | 0.001 | (0.001)** |
| Year fixed effects | Yes | | Yes | | Yes | | Yes | | Yes | |
| Country fixed effects | Yes | | Yes | | Yes | | Yes | | Yes | |
| Countries (observations) | 148 | (7,257) | 148 | (7,092) | 148 | (6,962) | 148 | (6,980) | 147 | (6,679) |
| $R^2$ ($R^2$ adjusted) | 0.870 | (0.866) | 0.870 | (0.866) | 0.870 | (0.866) | 0.871 | (0.867) | 0.871 | (0.867) |

**Panel B:** The effect of academic freedom (AF) on innovation output (patent applications) <u>considering different lags of the dependent variable.</u>

| Model | (1) | | (2) | | (3) | | (4) | | (5) | |
|---|---|---|---|---|---|---|---|---|---|---|
| **Dependent variable** | $\ln(\text{Patent applications}_{t+0})$ | | $\ln(\text{Patent applications}_{t+1})$ | | $\ln(\text{Patent applications}_{t+2})$ | | $\ln(\text{Patent applications}_{t+3})$ | | $\ln(\text{Patent applications}_{t+4})$ | |
| **Independent variables** | Coeff. | (SE) | Coeff. | (SE). | Coeff. | (SE) | Coeff. | (SE) | Coeff. | (SE) |
| $AF_t$ (z) | 0.335 | (0.032)*** | 0.343 | (0.032)*** | 0.345 | (0.032)*** | 0.336 | (0.032)*** | 0.327 | (0.033)*** |
| *Controls* | | | | | | | | | | |
| Year fixed effects | Yes | | Yes | | Yes | | Yes | | Yes | |
| Country fixed effects | Yes | | Yes | | Yes | | Yes | | Yes | |
| Countries (observations) | 157 | (12,706) | 157 | (12,549) | 157 | (12,392) | 157 | (12,235) | 157 | (12,078) |
| $R^2$ ($R^2$ adjusted) | 0.807 | (0.803) | 0.809 | (0.805) | 0.811 | (0.807) | 0.812 | (0.807) | 0.812 | (0.808) |

**Panel C:** The effect of academic freedom (AF) on innovation output (patent applications) <u>with the sample split according to the total number of patent applications per country.</u>

| Model | (1) | | (2) | | (3) | | (4) | | (5) | |
|---|---|---|---|---|---|---|---|---|---|---|
| **Sample** | Top 25 countries | | Top 50 countries | | Top 75 countries | | Countries not in top 75 | | Excluding countries with 0 patents | |
| **Dependent variable** | $\ln(\text{Patent applications}_{t+2})$ | | $\ln(\text{Patent applications}_{t+2})$ | | $\ln(\text{Patent applications}_{t+2})$ | | $\ln(\text{Patent applications}_{t+2})$ | | $\ln(\text{Patent applications}_{t+2})$ | |
| **Independent variables** | Coeff. | (SE) | Coeff. | (SE). | Coeff. | Coeff. | Coeff. | (SE) | Coeff. | (SE) |
| $AF_t$ (z) | 0.765 | (0.089)*** | 0.441 | (0.051)*** | 0.398 | (0.040)*** | -0.057 | (0.019)*** | 0.336 | (0.033)*** |
| *Controls* | | | | | | | | | | |
| Year fixed effects | Yes | | Yes | | Yes | | Yes | | Yes | |
| Country fixed effects | Yes | | Yes | | Yes | | Yes | | Yes | |
| Observations | 25 | (2,799) | 50 | (5,344) | 75 | (7,172) | 82 | (5,220) | 142 | (11,668) |
| $R^2$ ($R^2$ adjusted) | 0.683 | (0.666) | 0.759 | (0.751) | 0.781 | (0.775) | 0.491 | (0.471) | 0.807 | (0.803) |

**Panel D:** The effect of academic freedom (AF) on innovation output (patent applications) only considering years after 1980.

| Model | (1) | | (2) | | (3) | | (4) | | (5) | |
|---|---|---|---|---|---|---|---|---|---|---|
| **Dependent variable** | ln(Patent applications$_{t+2}$) | | ln(Patent applications$_{t+2}$) | | ln(Patent applications$_{t+2}$) | | ln(Patent applications$_{t+2}$) | | ln(Patent applications$_{t+2}$) | |
| **Independent variables** | Coeff. | (SE) | Coeff. | Coeff. | Coeff. | (SE) | Coeff. | (SE) | Coeff. | (SE) |
| AF$_t$ (z) | 0.311 | (0.060)*** | | | | | | | | |
| AF$_t$ (IV: Dem.$_{t-5}$, z) | | | 0.626 | (0.105)*** | | | | | | |
| AF$_t$ (IV: Dem. stock$_{t-10}$, z) | | | | | 0.682 | (0.096)*** | | | | |
| AF$_t$ (IV: AF$_{t-5}$, z) | | | | | | | 0.578 | (0.083)*** | | |
| AF$_t$ (IV: AF stock$_{t-10}$, z) | | | | | | | | | 0.567 | (0.079)*** |
| *Controls* | | | | | | | | | | |
| Year fixed effects | Yes | | Yes | | Yes | | Yes | | Yes | |
| Country fixed effects | Yes | | Yes | | Yes | | Yes | | Yes | |
| Observations | 157 | (5,190) | 157 | (5,042) | 157 | (4,924) | 157 | (5,046) | 156 | (4,892) |
| R² (R² adjusted) | 0.919 | (0.916) | 0.920 | (0.917) | 0.920 | (0.917) | 0.920 | (0.917) | 0.921 | (0.917) |

**Panel E:** The effect of academic freedom (AF) on innovation output (patent applications) <u>with an alternative *de jure* metric of AF</u>.

| Model | (1) | | (2) | | (3) | | (4) | | (5) | |
|---|---|---|---|---|---|---|---|---|---|---|
| **Dependent variable** | ln(Patent applications$_{t+2}$) | | ln(Patent applications$_{t+2}$) | | ln(Patent applications$_{t+2}$) | | ln(Patent applications$_{t+2}$) | | ln(Patent applications$_{t+2}$) | |
| **Independent variables** | Coeff. | (SE) | Coeff. | (SE) | Coeff. | (SE) | Coeff. | (SE) | Coeff. | (SE) |
| AF$_t$ | 0.520 | (0.060)*** | | | | | | | | |
| AF$_t$ (IV: Democracy$_{t-5}$) | | | 2.885 | (0.402)*** | | | | | | |
| AF$_t$ (IV: Democracy stock$_{t-10}$) | | | | | 2.849 | (0.390)*** | | | | |
| AF$_t$ (IV: AF$_{t-5}$) | | | | | | | 0.628 | (0.086)*** | | |
| AF$_t$ (IV: AF stock$_{t-10}$) | | | | | | | | | 0.611 | (0.086)*** |
| Year fixed-effects | Yes | | Yes | | Yes | | Yes | | Yes | |
| Country fixed effects | Yes | | Yes | | Yes | | Yes | | Yes | |
| Countries (observations) | 156 | (10,269) | 155 | (9,883) | 154 | (9,474) | 155 | (9,145) | 154 | (7,941) |
| R² (R² adjusted) | 0.825 | (0.820) | 0.794 | (0.788) | 0.798 | (0.792) | 0.832 | (0.827) | 0.839 | (0.834) |

*Notes*: Robust standard errors in parentheses. p < 0.10, ** p < 0.05, *** p < 0.01.

**Table 5:** Robustness tests for the effect of academic freedom (AF) on innovation quality (forward citations received in the first 3 years).

**Panel A:** The effect of academic freedom (AF) on innovation output (forward citations) <u>with additional control variables</u>.

| Model | (1) | | (2) | | (3) | | (4) | | (5) | |
|---|---|---|---|---|---|---|---|---|---|---|
| **Dependent variable** | ln(Forward citations$_t$+5) | | ln(Forward citations$_t$+5) | | ln(Forward citations$_t$+5) | | ln(Forward citations$_t$+5) | | ln(Forward citations$_t$+5) | |
| **Independent variables** | Coeff. | (SE) | Coeff. | Coeff. | Coeff. | (SE) | Coeff. | (SE) | Coeff. | (SE) |
| AF$_t$ (z) | 0.217 | (0.024)*** | | | | | | | | |
| AF$_t$ (IV: Dem.$_{t-5}$, z) | | | 0.457 | (0.057)*** | | | | | | |
| AF$_t$ (IV: Dem. Stock$_{t-10}$, z) | | | | | 0.466 | (0.057)*** | | | | |
| AF$_t$ (IV: AF$_{t-5}$, z) | | | | | | | 0.388 | (0.037)*** | | |
| AF$_t$ (IV: AF stock$_{t-10}$, z) | | | | | | | | | 0.418 | (0.037)*** |
| *Controls* | | | | | | | | | | |
| ln(Patent applications) | 0.250 | (0.016)*** | 0.249 | (0.016)*** | 0.250 | (0.016)*** | 0.253 | (0.016)*** | 0.257 | (0.016)*** |
| GDP per capita | 0.022 | (0.003)*** | 0.025 | (0.003)*** | 0.028 | (0.003)*** | 0.030 | (0.003)*** | 0.035 | (0.004)*** |
| Ln(Population) | -0.833 | (0.068)*** | -0.749 | (0.081)*** | -0.746 | (0.082)*** | -0.806 | (0.075)*** | -0.845 | (0.082)*** |
| Migration rate | -0.001 | (0.001)** | -0.001 | (0.001)* | -0.001 | (0.001) | -0.001 | (0.001) | -0.001 | (0.001) |
| Year fixed effects | Yes | | Yes | | Yes | | Yes | | Yes | |
| Country fixed effects | Yes | | Yes | | Yes | | Yes | | Yes | |
| Countries (observations) | 148 | (6,816) | 148 | (6,651) | 148 | (6,521) | 147 | (6,540) | 147 | (6,241) |
| R² (R² adjusted) | 0.892 | (0.889) | 0.891 | (0.888) | 0.892 | (0.888) | 0.893 | (0.890) | 0.894 | (0.890) |

**Panel B:** The effect of academic freedom (AF) on innovation output (forward citations) <u>considering different lags of the dependent variable</u>.

| Model | (1) | | (2) | | (3) | | (4) | | (5) | |
|---|---|---|---|---|---|---|---|---|---|---|
| **Dependent variable** | ln(Forward citations$_t$+0) | | ln(Forward citations$_t$+1) | | ln(Forward citations$_t$+2) | | ln(Forward citations$_t$+3) | | ln(Forward citations$_t$+4) | |
| **Independent variables** | Coeff. | (SE) | Coeff. | (SE). | Coeff. | (SE) | Coeff. | (SE) | Coeff. | (SE) |
| AF$_t$ (z) | 0.174 | (0.022)*** | 0.192 | (0.022)*** | 0.210 | (0.022)*** | 0.228 | (0.022)*** | 0.245 | (0.022)*** |
| *Controls* | | | | | | | | | | |
| ln(Patent applications) | Yes | | Yes | | Yes | | Yes | | Yes | |
| Year fixed effects | Yes | | Yes | | Yes | | Yes | | Yes | |
| Country fixed effects | Yes | | Yes | | Yes | | Yes | | Yes | |
| Countries (observations) | 157 | (12,706) | 157 | (12,549) | 157 | (12,392) | 157 | (12,235) | 157 | (12,078) |
| R² (R² adjusted) | 0.616 | (0.608) | 0.621 | (0.613) | 0.626 | (0.618) | 0.631 | (0.623) | 0.636 | (0.628) |

**Panel C:** The effect of academic freedom (AF) on innovation output (forward citations) <u>with the sample split according to the total number of forward citations per country</u>.

| Model | (1) | | (2) | | (3) | | (4) | |
|---|---|---|---|---|---|---|---|---|
| **Sample** | Top 25 countries | | Top 50 countries | | Countries not in top 50 | | Excluding countries with 0 patents | |
| **Dependent variable** | ln(Forward citations$_t$+5) | | ln(Forward citations$_t$+5) | | ln(Forward citations$_t$+5) | | ln(Forward citations$_t$+5) | |
| **Independent variables** | Coeff. | (SE) | Coeff. | (SE). | Coeff. | Coeff. | Coeff. | (SE) |
| AF$_t$ (z) | 0.399 | (0.063)*** | 0.362 | (0.042)*** | -0.001 | (0.002) | 0.395 | (0.032)*** |
| *Controls* | | | | | | | | |
| ln(Patent applications) | Yes | | Yes | | Yes | | Yes | |
| Year fixed effects | Yes | | Yes | | Yes | | Yes | |
| Country fixed effects | Yes | | Yes | | Yes | | Yes | |
| Observations | 25 | (2,664) | 51 | (4,943) | 106 | (6,978) | 73 | (6,576) |
| R² (R² adjusted) | 0.756 | (0.743) | 0.681 | (0.670) | 0.165 | (0.138) | 0.653 | (0.643) |

**Panel D:** The effect of academic freedom (AF) on innovation output (forward citations) only considering years after 1980.

| Model | (1) | | (2) | | (3) | | (4) | | (5) | |
|---|---|---|---|---|---|---|---|---|---|---|
| **Dependent variable** | ln(Forward citations$_t$+5) | | ln(Forward citations$_t$+5) | | ln(Forward citations$_t$+5) | | ln(Forward citations$_t$+5) | | ln(Forward citations$_t$+5) | |
| **Independent variables** | Coeff. | (SE) | Coeff. | Coeff. | Coeff. | (SE) | Coeff. | (SE) | Coeff. | (SE) |
| AF$_t$ (z) | 0.137 | (0.033)*** | | | | | | | | |
| AF$_t$ (IV: Dem.$_{t-5}$, z) | | | 0.637 | (0.086)*** | | | | | | |
| AF$_t$ (IV: Dem. stock$_{t-10}$, z) | | | | | 0.650 | (0.083)*** | | | | |
| AF$_t$ (IV: AF$_{t-5}$, z) | | | | | | | 0.335 | (0.052)*** | | |
| AF$_t$ (IV: AF stock$_{t-10}$, z) | | | | | | | | | 0.372 | (0.051)*** |
| *Controls* | | | | | | | | | | |
| ln(Patent applications) | Yes | | Yes | | Yes | | Yes | | Yes | |
| Year fixed effects | Yes | | Yes | | Yes | | Yes | | Yes | |
| Country fixed effects | Yes | | Yes | | Yes | | Yes | | Yes | |
| Observations | 157 | (4,719) | 157 | (4,571) | 157 | (4,453) | 156 | (4,576) | 156 | (4,424) |
| R² (R² adjusted) | 0.913 | (0.909) | 0.907 | (0.903) | 0.907 | (0.903) | 0.913 | (0.909) | 0.912 | (0.909) |

**Panel E:** The effect of academic freedom (AF) on innovation output (forward citations) <u>with an alternative *de jure* metric of AF</u>.

| Model | (1) | | (2) | | (3) | | (4) | | (5) | |
|---|---|---|---|---|---|---|---|---|---|---|
| **Dependent variable** | ln(Forward citations$_t$+5) | | ln(Forward citations$_t$+5) | | ln(Forward citations$_t$+5) | | ln(Forward citations$_t$+5) | | ln(Forward citations$_t$+5) | |
| **Independent variables** | Coeff. | (SE) | Coeff. | (SE) | Coeff. | (SE) | Coeff. | (SE) | Coeff. | (SE) |
| AF$_t$ | 0.280 | (0.042)*** | | | | | | | | |
| AF$_t$ (IV: Democracy$_{t-5}$) | | | 6.990 | (0.566)*** | | | | | | |
| AF$_t$ (IV: Democracy stock$_{t-10}$) | | | | | 7.028 | (0.538)*** | | | | |
| AF$_t$ (IV: AF$_{t-5}$) | | | | | | | 0.461 | (0.065)*** | | |
| AF$_t$ (IV: AF stock$_{t-10}$) | | | | | | | | | 0.391 | (0.068)*** |
| *Controls* | | | | | | | | | | |
| ln(Patent applications) | Yes | | Yes | | Yes | | Yes | | Yes | |
| Year fixed-effects | Yes | | Yes | | Yes | | Yes | | Yes | |
| Country fixed effects | Yes | | Yes | | Yes | | Yes | | Yes | |
| Countries (observations) | 155 | (9,819) | 154 | (9,433) | 153 | (9,024) | 155 | (8,696) | 153 | (7,500) |
| R² (R² adjusted) | 0.658 | (0.648) | - | | - | | 0.679 | (0.669) | 0.705 | (0.695) |

*Notes*: Robust standard errors in parentheses. $p < 0.10$, ** $p < 0.05$, *** $p < 0.01$.